\documentclass[a4paper,12pt]{article}
\usepackage{amssymb}
\usepackage{amsmath}
\usepackage{amsfonts}
\usepackage{amsthm}
\usepackage{graphicx}
\usepackage{pst-all}
\begin{document}

\newcommand{\aka}{\ensuremath{a^\dag a}}
\newcommand{\raka}{\ensuremath{\overrightarrow{a^\dag a}}}
\newcommand{\laka}{\ensuremath{\overleftarrow{a^\dag a}}}
\newcommand{\ak}{\ensuremath{a^\dag}}
\newcommand{\rak}{\ensuremath{\overrightarrow{a}^\dag}}
\newcommand{\lak}{\ensuremath{\overleftarrow{a}^\dag}}
\newcommand{\ra}{\ensuremath{\overrightarrow{a}}}
\newcommand{\la}{\ensuremath{\overleftarrow{a}}}
\newcommand{\bkb}{\ensuremath{b^\dag b}}
\newcommand{\bk}{\ensuremath{b^\dag }}
\newcommand{\rbk}{\ensuremath{\overrightarrow{b}^\dag}}
\newcommand{\lbk}{\ensuremath{\overleftarrow{b}^\dag}}
\newcommand{\rb}{\ensuremath{\overrightarrow{b}}}
\newcommand{\lb}{\ensuremath{\overleftarrow{b}}}
\newcommand{\E}{\ensuremath{\mathbb E}}
\newcommand{\N}{\ensuremath{\mathbb {N}_t}}
\newcommand{\ket}{\ensuremath{\rangle}}
\newcommand{\bra}{\ensuremath{\langle}}
\newcommand{\tr}{\ensuremath{\mathrm {Tr}}}
\newcommand{\im}{\ensuremath{\Im\mathrm {m}}}
\newcommand{\re}{\ensuremath{\Re\mathrm {e}}}
\newcommand{\ar}[1]{\ensuremath{\overrightarrow{#1}}}
\newcommand{\al}[1]{\ensuremath{\overleftarrow{#1}}}
\renewcommand{\atop}[2]{\genfrac{}{}{0pt}{}{#1}{#2}}
\newcommand{\osc}{\mathrm{osc}}

\title{On a quantum model of a laser-interferometer measuring a weak classical force}
\author {A.~M.~Sinev}
\date{}
\maketitle

\begin{abstract}
We consider a solvable model of a laser-interferometer measuring a
weak classical force. The model takes into account dissipation of
the energy by transfer to the environment at zero temperature. The
sensitivity (the signal-to-noise ratio) of the device is defined as
the corresponding ratio between the mean value  and the variance of
a certain observable. We analyze the dependence of the sensitivity
upon the duration of the measurement and the photon number. For
parameters typical for the LIGO project, we discuss numerical
estimates.
\end{abstract}

\section{Introduction}
In this paper, we consider the measurement of a weak classical force
perturbed by quantum effects. The measuring device consists of a
suspended mirror considered as a quantum oscillator driven by a
classical force, and a recording device which produces the reduction
of the sensor state.

The coherent electromagnetic field, considered as a sensor, is
enclosed in the cavity between the movable and the fixed mirrors.
Any displacement of the oscillator changes the phase of the wave
leaving the cavity. The phase is measured by the interferometer, and
the output signal provides an information about the classical force.
We assume that the interaction between the oscillator and the
environment is irreversible: dissipation of the energy is the result
of spontaneous transfer of internal mechanical tensions in the
suspension of the mirror to acoustic waves. By this reason, we treat
the oscillator as an open quantum system and its states as solutions
of the master Markov equation \cite{carmic}.

Under sufficiently general assumptions, the evolving state obeys the
Lindblad equation \cite{lin}
\begin{equation} \label{lin}
 \frac d{dt}\rho_t = \mathcal L^\dagger(\rho_t)\,,\qquad
 \mathcal L^\dagger (\rho_t) = \Phi^\dag (\rho_t) - \Phi(I)\circ
 \rho_t - \frac i\hbar[\widehat{H}, \rho_t]\,,
\end{equation}
where $\widehat{H}$ is the system Hamiltonian describing the reversible
dynamics, $\Phi(\cdot)$ is a certain completely positive map describing
dissipative processes, $\Phi^\dag(\cdot)$
is the completely positive map dual to $\Phi(\cdot)$. The duality is established by
the trace:
$$
\tr\,\Phi^\dagger(\rho)\sigma=\tr\,\rho\Phi(\sigma).
$$
$A \circ B = \frac12(AB+BA)$ is the symmetric Jordan product.

The specific form of the map $\mathcal L^\dagger$ can be deduced
from a physical model of a coupling between the system and the
environment \cite{bgar}. For example, the intensity of the
dissipative transfer of the energy is proportional to the energy of
the state, and the evolution of any state to the ground state (zero
temperature) is described by a CCP Lindblad generator:
\begin{equation}\label{damp}
{\cal L}_0^\dagger(\rho_t)= \lambda\left(
b\rho_t\bk-\frac{1}{2}(\rho_t\bkb+ \bkb\rho_t) \right)\, ,
\end{equation}
where $\lambda>0$ is the dissipation rate, i.e. the mean number of
quantums of energy transferred per unit of time to the environment,
$\bk\text{ and }b$ are the creation and annihilation operators,
$\bkb$ is the energy of the oscillator.

If the reservoir has a nonzero temperature, the generator driving
the compound system to the equilibrium has the form \cite{gar,eng}:
\begin{equation} \label{hibbs}
{\cal L}_\nu^\dagger(\rho_t) = \frac{\lambda}2 (\nu+1)( 2b\rho_t\bk
-\bk b \rho_t-\rho_t \bk b )+\frac{\lambda}2 \nu( 2\bk\rho_t b -b\bk
\rho_t-\rho_t b\bk)\,.
\end{equation}
The mean energy of the quantum oscillator with eigenfrequency
$\Omega$ in the balanced state with such environment is equal to
$E_0 = \hbar\Omega \nu$.

This paper extends our previous work \cite{sin}. In the first
section, we describe a mathematical model of the oscillator
interacting with the laser radiation and a classical force, but
unlike \cite{sin}, we take into account the transfer of the energy
to the environment at zero temperature. For the product system
consisting of the oscillator and the electromagnetic field, we
derive an explicit form of the density matrix.

This solution is applied to calculation of the mean value of the
intensity and the variance of varying photon flow at the output of a
twin-wave interferometer of a Michelson or Mach-Zehnder type. In
comparison with \cite{sin}, here we use an observable which properly
takes into account the two-beam quantum interference.

In the third section, we consider relative fluctuations of the
signal. Numerical estimates of the sensitivity are based on
realistic data describing the LIGO detector. Physically motivated
estimates of the sensitivity of such devices based on spectral
representation can be found in \cite{bra1}, \cite{bra}.

\section{Solvable model}
Consider the coherent electromagnetic field with frequency $\omega$
between a movable mirror and a fixed one. Let a small classical
external force act on the movable mirror. By $\ak\text{ and }a $ we
denote the creation and annihilation operators of the radiation in
the cavity, and $\bk\text{ and } b$ stand for the creation and
annihilation operators of the quantum oscillator (the movable
mirror) with eigenfrequency  $\Omega$.

The reversible dynamics of the laser radiation and the oscillator in
the Hilbert space $l_2\otimes l_2$ (the first factor corresponds to
the space of radiation states, and the second one corresponds to
that of the oscillator) is described by the generator consisting of
three summands \cite{bos, brif}
\begin{equation} \label{Ham}
H_t=\hbar\left[\omega\aka\otimes I+I\otimes\Omega \bk
b+(g\,\aka+f(t))\otimes(\bk+b)\right],
\end{equation}
where the first two terms are the energy operators of the radiation
and the oscillator, and the third one is the energy of the
oscillator due to the radiation pressure and an external force,
$\hat{x}=\frac{\bk +b}{\sqrt{2}}$ is the position operator of the
movable mirror. We used the following notation:
\begin{equation}\label{constg}
g = \frac{\omega}{L}\sqrt{\frac{2\hbar}{m\Omega}}
\end{equation}
is the constant of coupling  between the laser radiation and the oscillator,
\begin{equation}\label{constf}
f(t)=\frac{F(t)}{\sqrt{2m\Omega\hbar}},
\end{equation}
$F(t)$ is a classical force, $m$ is the mass of the mirror, $L$ is
the distance between the mirrors. In accordance with \eqref{lin} and
\eqref{damp}, the evolution of the system with  Hamiltonian
\eqref{Ham} and dissipation of the energy to the environment at zero
temperature is described  by a quantum master equation
\begin{equation} \label{eq}
\frac{d}{dt}\rho_t=-\frac i \hbar[H_t,\rho_t] -\frac{\lambda}2 (\bk
b \rho_t+\rho_t \bk b -2b\rho_t\bk), \quad
\rho_t\bigl|_{t=0}=\rho_0.
\end{equation}

We look for the solution of this equation in the following form
\cite{cheb}:
\begin{equation}\label{qt}
\rho_t=\E \,u_t \rho_0 u_t^* \,,
\end{equation}
where $u_t$ is an operator-valued stochastic function satisfying the
stochastic Schr\"{o}dinger equation
\begin{equation} \label{stoch}
du_t=\left[\left(-\frac{\lambda}{2}\bkb-i\Omega\,\bkb-i\omega
\aka-i\left(g\,\aka+f(t)\right)(\bk+b)\right)dt+
\sqrt{\lambda}\,b\,dw_t\right]u_t,
\end{equation}
with initial condition $u_0 = I$. $w_t$ stands for the standard
Wiener process. The representation (\ref{qt}), (\ref{stoch}) for the
solution of (\ref{eq}) is well known (\cite{bgar}, \cite{gar}); it
follows from the Ito differentiation rule \cite{ito}.

Passing to the interaction representation \cite{sin}, setting $ v_t
\stackrel{\text{\textit{def}}}{=} e^{i \omega \aka \,t}e^{(i\Omega
+\frac\lambda2)\bkb \,t}u_t$, we have
\[
d\, v_t = \left[-i\bigl(g\aka+f(t)\bigr)(b e^{-(i\Omega
+\frac\lambda2)t} + \bk e^{(i\Omega +\frac\lambda2)t})dt +
\sqrt\lambda b e^{-(i\Omega +\frac\lambda2)t}dw_t \right]v_t.
\]
Then, $\phi_t \stackrel{\text{\textit{def}}}{=}
e^{i\bk\beta^+_t}v_t$, so we obtain
\begin{gather*}
d\phi_t = \left[-i(g\aka + f(t))(b-i\beta^+_t)e^{-(i\Omega
+\frac\lambda2)t})dt +\sqrt\lambda (b -i\beta^+_t ) e^{-(i\Omega
+\frac\lambda2)t}dw_t\right]\phi_t.
\end{gather*}
In the above equation, all summands commute. The notation
$\beta^{\pm}_t$ is used for the following family of commuting operators:
\begin{equation} \label{beta}
\beta^{\pm}_t = \int\limits_0^t(g\aka + f(\tau)) e^{\pm(i\Omega
+\frac\lambda2)\tau}d\tau.
\end{equation}

Picking out a stochastic part of the solution, we obtain an equation
for $\xi_t \stackrel{def}{=} e^{ib\beta^-_t}e^{C_t}\phi_t$:
\begin{equation}
d\xi_t = \sqrt\lambda (b -i\beta^+_t ) e^{-(i\Omega
+\frac\lambda2)t}\xi_tdw_t, \quad C_t = \int\limits_0^t(g\aka +
f(\tau))e^{-(i\Omega +\frac\lambda2)\tau} \beta^+_\tau
d\tau.\label{c}
\end{equation}
The solution $\xi_t$ is a stochastic process which has the form of
the operator-valued Girsanov functional \cite{Gi60}:
\[
\xi_t = e^{\sqrt\lambda\int_0^t(b -i\beta^+_\tau ) e^{-(i\Omega
+\frac\lambda2)\tau}dw_\tau} e^{-\frac\lambda2\int_0^t(b
-i\beta^+_\tau )^2 e^{-2(i\Omega +\frac\lambda2)\tau}d\tau}.
\]
Finally, the solution of equation \eqref{stoch} reads as the
following normally ordered composition of exponents:
\begin{align}
u_t &= e^{-i \omega \aka \,t}e^{-(i\Omega +\frac\lambda2)\bkb \,t}
e^{-i\bk\beta^+_t} e^{-ib\beta^-_t}e^{-C_t} \notag\\
&\times e^{\sqrt\lambda\int_0^t(b -i\beta^+_\tau ) e^{-(i\Omega
+\frac\lambda2)\tau}dw_\tau} e^{-\frac\lambda2\int_0^t(b
-i\beta^+_\tau )^2 e^{-2(i\Omega
+\frac\lambda2)\tau}d\tau}.\label{u}
\end{align}

Consider the solution of problem \eqref{eq} according to \eqref{qt}.
Taking nonrandom factors out of the mathematical expectation, we
obtain
\begin{align}\label{rho}
\rho_t &=e^{-i \omega \aka \,t}\,e^{-(i\Omega +\frac\lambda2)\bkb
\,t}\, e^{-i\bk\beta^+_t}\, e^{-ib\beta^-_t}\,e^{-C_t}
\,e^{-\frac\lambda2\int_0^t(b -i\beta^+_\tau )^2
e^{-2(i\Omega +\frac\lambda2)\tau}d\tau}\notag\\
&\qquad\times \E\, \biggl(e^{\sqrt\lambda\int_0^t(b -i\beta^+_\tau )
e^{-(i\Omega +\frac\lambda2)\tau}dw_\tau}\,\rho_0\,
 e^{\sqrt\lambda\int_0^t(\bk +i\beta^{+*}_\tau )
e^{(i\Omega -\frac\lambda2)\tau}dw_\tau}\biggr)
\notag\\
&\qquad\times \,e^{-\frac\lambda2\int_0^t(\bk +i\beta^{+*}_\tau )^2
e^{2(i\Omega -\frac\lambda2)\tau}d\tau}\, e^{-C_t^*}\,
e^{i\bk\beta^{-*}_t}\, e^{i b\beta^{+*}_t}\,e^{(i\Omega
-\frac\lambda2)\bkb \,t}\,e^{i
\omega \aka \,t} \notag\\
 &= e^{-i \omega \aka \,t}\,e^{-(i\Omega
+\frac\lambda2)\bkb \,t}\, e^{-i\bk\beta^+_t}\,
e^{-ib\beta^-_t}\,e^{-C_t} \,
e^{ib\lambda\int_0^t\al{\beta}^{+*}_\tau e^{-\lambda\tau}d\tau}\notag\\
&\qquad\times \biggl(e^{\ar{b}\al{\bk}(1-e^{-\lambda
t})}\,e^{\lambda\int_0^t\ar{\beta}^+_\tau \al{\beta}^{+*}_\tau
e^{-\lambda\tau}d\tau}\,\rho_0\biggr)\notag \\
&\qquad\times e^{i\bk\lambda\int_0^t\ar{\beta}^+_\tau
e^{-\lambda\tau}d\tau} \, e^{-C_t^*}\, e^{i\bk\beta^{-*}_t}\, e^{i
b\beta^{+*}_t}\,e^{(i\Omega -\frac\lambda2)\bkb \,t}\,e^{i \omega
\aka \,t}\,.
\end{align}
The mathematical expectation \eqref{rho} is evaluated explicitly
because the stochastic processes represented as Ito integrals over the
Wiener measure in the exponents of \eqref{u} are the Gaussian processes.
The arrows indicate the ordering of operators with respect to $\rho_0$.

Let us find the mean energy of the oscillator $\bra \bkb\ket$.
Suppose that the oscillator is prepared in the ground state and the
initial state of the  radiation is the coherent state with mean
photon number $N=|z|^2$: $\rho_0=|0\ket\bra0|_{osc}\otimes|z\ket\bra
z|_{las}$. Therefore,
\begin{align}
\bra \bkb\ket
&=\tr\left\{\bkb\,\rho_0\right\}=e^{-|z|^2}\sum_{n}\frac{|z|^2}{n!}\,
e^{\lambda\int_0^t\beta^+_n(\tau) \beta^{+*}_n(\tau)\,
e^{-\lambda\tau}d\tau}\,e^{-C_n}\,e^{-C_n^*}\notag\\
&\times\bra 0| e^{i b\,e^{(i\Omega-\frac\lambda2)t}
\beta^{+*}_n}\bkb \,e^{-i\bk
e^{-(i\Omega+\frac\lambda2)t}\beta^+_n}|0\ket
=e^{-|z|^2}\sum_{n}\frac{|z|^2}{n!}|\beta_n^+|^2 e^{-\lambda t}.
\end{align}
By using the equality $\beta_n^+= \int\limits_0^t\bigl(g n+
f(\tau)\bigr) e^{(i\Omega +\frac\lambda2)\tau}d\tau$, we find that
the mean energy consists of three summands
\[
\bra \bkb\ket = g^2(N^2+N)c_2(t)+gNc_1(t)+c_0(t)\,,\quad N=|z|^2,
\]
where
\begin{gather}
c_2 =\frac{1-2e^{-\frac{\lambda t}2}\cos\Omega t+e^{-\lambda
t}}{\Omega^2+\frac{\lambda^2}4}, \\
c_1=2\left[\frac{\sin (\Omega t+\Phi)-\sin\Phi\,e^{-\frac{\lambda
t}2}}{\sqrt{\Omega^2+\frac{\lambda^2}4}} \int_0^tf(\tau)
e^{\frac\lambda2(\tau-t)}\cos\Omega \tau\,d\tau\right.\notag\\
\left.-\frac{\cos (\Omega t+\Phi)-\cos\Phi\,e^{-\frac{\lambda
t}2}}{\sqrt{\Omega^2+\frac{\lambda^2}4}} \int_0^tf(\tau)
e^{\frac\lambda2(\tau-t)}\sin\Omega \tau\,d\tau\right],\;\tan\Phi=\frac\lambda{2\Omega},\\
c_0=e^{-\lambda
t}\left|\int_0^tf(\tau)e^{(i\Omega+\frac\lambda2)\tau}d\tau\right|^2.
\end{gather}
In particular, if $f(t)\equiv0$ and $N\gg1$, we have
\begin{equation}\label{meanen}
\bra \bkb\ket = g^2N^2c_2(t)
\stackrel{t\to\infty}{=}\frac{g^2N^2}{\Omega^2+\frac{\lambda^2}4}.
\end{equation}
This quantity corresponds to the classical shift of the coordinate
caused by the light pressure.

Indeed, during the time $t = \frac {2L}c$, where $L$ is the distance
between the mirrors, all $N$ photons of the cavity are reflected
from the movable mirror. The momentum transferred to the mirror
amounts to $p=2\frac{N\hbar\omega}c$. Consequently, the
pressure on the oscillator equals  $F= \frac pt =
\frac{N\hbar\omega}L$. For a stationary state, $F=m\Omega^2 x$. Hence,
$x=\frac{N\hbar\omega}{Lm\Omega^2}$, and the total energy equals
\begin{equation}\label{encl}
E=\frac{m\Omega^2x^2}2 =\frac{N^2\hbar^2\omega^2}{2L^2m\Omega^2}.
\end{equation}
Taking $g=\frac\omega L\sqrt{\frac{\hbar}{2m\Omega}}$ as a coupling constant,
we obtain
\[
\hbar\Omega\bra \bkb\ket
=\frac{N^2\hbar^2\omega^2}{2L^2m(\Omega^2+\frac{\lambda^2}4)}.
\]
It differs from $E$ in \eqref{encl} by the summand
$\frac{\lambda^2}4$ in the denominator. Thus, the dissipation
of the energy in this model is proportional to the total energy of the
oscillator.
\section{Interferometric measurement of a phase shift}
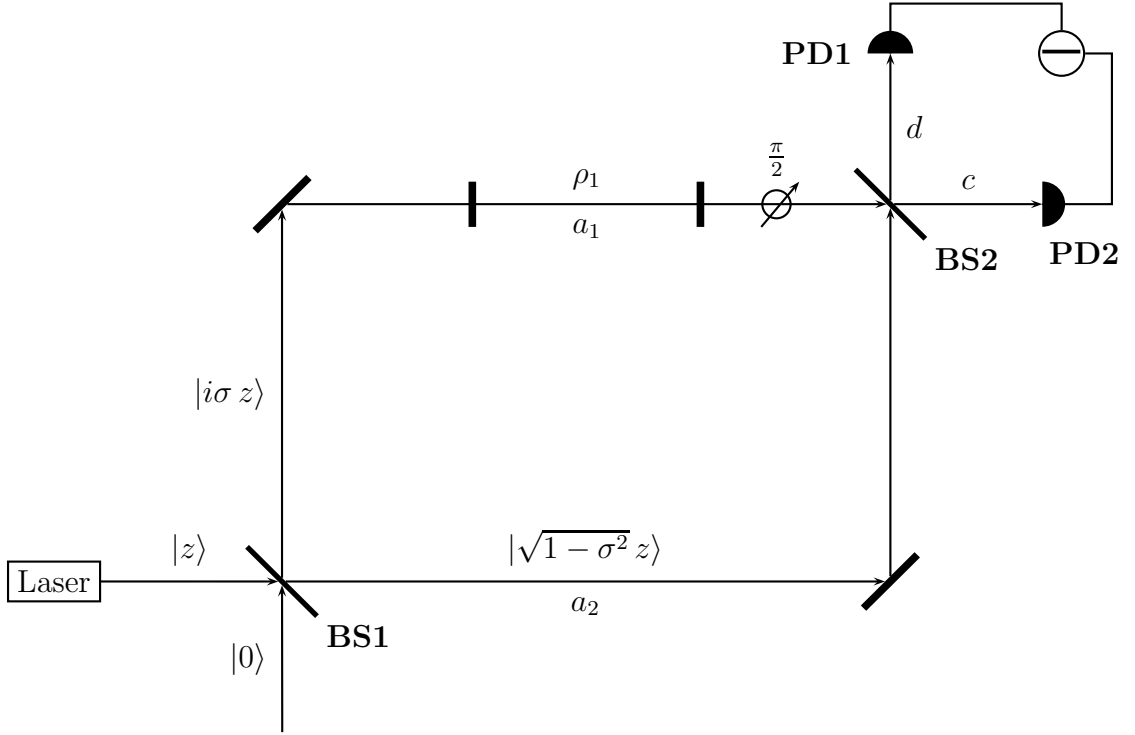
\begin{figure}
%\begin{center}
\begin{pspicture}(-1,0)(14,8)
\rput(0,2){\rnode{las}{\psframebox{Laser}}}%
\rput{45}(3,2){\fnode*[framesize=.07 1.3]{bs1}}%
\nput{-45}{bs1}{\textbf{BS1}}%
\pnode(3,0){vac}%
\ncline{->}{las}{bs1}%
\naput{$|z\ket$}%
\ncline{->}{vac}{bs1}%
\naput{$|0\ket$}%
\rput{135}(11,2){\fnode*[framesize=.1 1]{mir1}}%
\ncline{->}{bs1}{mir1}%
\naput{$|\sqrt{1-\sigma^2}\,z\ket$}%
\nbput{$a_2$}%
\rput{-45}(3,7){\fnode*[framesize=.1 1]{mir2}}%
\ncline{->}{bs1}{mir2}%
\naput{$|i\sigma\,z\ket$}%
\rput{45}(11,7){\fnode*[framesize=.07 1.3]{bs2}}%
\nput{-45}{bs2}{\textbf{BS2}}%
\ncline{->}{mir2}{bs2}%
\naput{$\rho_1$} \nbput{$a_1$}%
\ncline{->}{mir1}{bs2}%
\psline[linewidth=0.1](5.5,7.3)(5.5,6.7)
\psline[linewidth=0.1](8.5,7.3)(8.5,6.7)
\rput(11,9){\rnode{pd1}{\pswedge*{0.3}{0}{180}}}%
\nput[labelsep=0.5]{180}{pd1}{\textbf{PD1}}%
\ncline{->}{bs2}{pd1} \nbput{$d$}%
\rput(13,7){\rnode{pd2}{\pswedge*{0.3}{-90}{90}}}%
\nput[labelsep=0.5]{-60}{pd2}{\textbf{PD2}}%
\ncline{->}{bs2}{pd2}  \naput{$c$}%
\rput(13,9){\psframe(0.5,0.05)} \rput(13.25,9){\cnode{0.31}{min}}
\ncbar{pd2}{min}
\ncbar[angle=90]{pd1}{min}%
\cnode(9.5,7){0.2}{ph}%
\nput{90}{ph}{$\frac\pi2$}%
\psline{->}(9.3,6.7)(9.8,7.3)
\end{pspicture}
%\end{center}
\caption{The beam splitting in the two-arm interferometer}\label{int}
\end{figure}

Consider the laser beam splitting scheme in the two-arm interferometer (Fig. \ref{int}).
The  coherent electromagnetic radiation falls on the beam splitter
(BS1) with reflectivity $\sigma$ and  transmissivity $\sqrt
{1-\sigma^2}$. The movable mirror interacts with the reflected beam.
Passing through the cavity, the signal wave carries system
information and interferes with the carrier wave on the second beam
splitter (BS2) with a splitting ratio 50/50. Preliminary, the phase
of the signal wave is shifted  by $\frac \pi 2$. The interferometer
output contains the balanced detector with two photo detectors (PD1,
PD2) which measure the intensity of the interferenced light.

Let the unitary scattering matrices of the beam splitters be chosen as
\cite{opt}
\begin{equation}\label{split}
\begin{bmatrix}
\sqrt{1-\sigma^2}& i\sigma\\
i\sigma&\sqrt{1-\sigma^2}
\end{bmatrix}, \quad |\sigma|\le 1\qquad
\text{and}\qquad \frac1{\sqrt{2}}
\begin{bmatrix}
1& i\\
i&1
\end{bmatrix}.
\end{equation}
Suppose that the radiation is in the coherent state
\[
|\psi\ket=|z\ket=e^{-\frac{|z|^2}2}(1, z,
\frac{z^2}{\sqrt{2!}},\dots, \frac{z^n}{\sqrt{n!}},\dots)
\]
with  amplitude  $z\in \mathbb C$ and mean photon number
$N=|z|^2$. According to
\eqref{split}, after passing through the first splitter,
 the states of reflected and transmitted beams are equal to
\begin{equation}
|\psi\ket_1=|i\sigma z\ket\qquad \text{and} \qquad
|\psi\ket_2=|\sqrt{1-\sigma^2}\,z\ket
\end{equation}
respectively. Hence, $|\psi\ket_1$ is the initial state of the
radiation inside the cavity.

The evolution of the radiation in the first arm is described by the
density operator $\rho_t$ \eqref{rho} averaged by the partial trace
$\tr_{\osc}$ over the states of the oscillator. Suppose that the
oscillator is prepared in the ground state\footnote{In what follows,
the steady state is independent of the initial state of the system,
so our choice $\rho_1(0)=|0\ket_{\osc}\! \phantom{i}_{\osc}\bra 0|$
does not cause any loss of generality.} $\rho_1(0)=|0\ket_{\osc}\!
\phantom{i}_{\osc}\bra 0|$. Therefore,
\begin{multline}\label{rhoint}
\rho_1(t) =  \tr_{\osc} \left\{e^{-i \omega \aka \,t}\,e^{-(i\Omega
+\frac\lambda2)\bkb \,t}\, e^{-i\bk\beta^+_t}\, \,e^{-C_t} \,
e^{\lambda\int_0^t\ar{\beta}^+_\tau \al{\beta}^{+*}_\tau
e^{-\lambda\tau}d\tau}\,|i\sigma z\ket_1
\phantom{i}_1\! \bra i\sigma z|\right. \\
\left.\otimes |0\ket_{\osc}\! \phantom{i}_{\osc}\bra 0|
\,e^{-C_t^*}\, e^{i b\beta^{+*}_t}\,e^{(i\Omega -\frac\lambda2)\bkb
\,t}\,e^{i \omega \aka \,t}\right\}\\
=e^{-i\omega(\overrightarrow{\aka}-\overleftarrow{\aka})t}e^{-\overrightarrow{C_t}-\overleftarrow{C_t^*}}
e^{\lambda\int_0^t\overrightarrow{\beta_\tau^+}
(\overleftarrow{\beta_\tau^+})^* e^{-\lambda\,\tau}d\tau}\,
e^{\overrightarrow{\beta_t^+} (\overleftarrow{\beta_t^+})^*
e^{-\lambda\,t}}|i\sigma z\ket_1 \phantom{i}_1\! \bra i\sigma z|\,.
\end{multline}
Using the definitions of $\beta_{t}^+\text{ and }C_t$ \eqref{beta},
\eqref{c}, we simplify expression \eqref{rhoint}
\begin{equation}
\rho_1(t) = e^{g(\laka-\raka)\int_0^t\int_0^\tau d\tau
ds\,[(g\raka-f(s))e^{(\frac\lambda2+i\Omega)(s-\tau)}-(g\laka-f(s))e^{(\frac\lambda2-i\Omega)(s-\tau)}]}
 |i\sigma z\ket_1 \phantom{i}_1\! \bra i\sigma z|\,.
\end{equation}
It is convenient to express the density matrix in terms of the
canonical basis
$|n\ket=\bigl(\underbrace{0,\dots,0,}_{n}1,0\dots\bigr)$
\begin{multline} \label{rhol}
\rho_1(t) = e^{-N \sigma^2}\sum_{n=0}^\infty \sum_{m=0}^\infty
|n\ket_1 \phantom{i}_1\bra m| \frac{(i\sigma z)^n (-i\sigma z^*)^m}{\sqrt{n!m!}} e^{i\omega(m-n)t}\\
\times e^{g^2(m-n)\int_0^t\int_0^\tau d\tau ds \,\left(n \,e^{(i
\Omega +\frac\lambda2)(s-\tau)}-m \,e^{(-i \Omega
+\frac\lambda2)(s-\tau)}\right)} \\ \times e^{2 i
g(m-n)\int_0^t\int_0^\tau d\tau ds \,f(s)
e^{\frac\lambda2(s-\tau)}\sin\Omega(s-\tau)}.
\end{multline}
The evolution of the radiation state in the reference arm of
interferometer (carrier wave) is given by the following equation:
\begin{align}\label{rho2}
\rho_2(t) &= e^{-i \ak_2 a^{\phantom{\dagger}}_2 \omega
t}|\sqrt{1-\sigma^2}z\ket_2 \phantom{i}_2\! \bra\sqrt{1-\sigma^2}z|
e^{i \ak_2 a^{\phantom{\dagger}}_2 \omega
t}\notag\\
&=e^{-N (1-\sigma^2)}\sum_{n=0}^\infty \sum_{m=0}^\infty |n\ket_2
\phantom{i}_2\bra m|\frac{(\sqrt{1-\sigma^2}z)^n
(\sqrt{1-\sigma^2}z^*)^m}{\sqrt{n!m!}} e^{i\omega(m-n)t}\,.
\end{align}

Finally, the output operator
of the balanced detector is equal to
\begin{equation}\label{charge}
\widehat{I}= c^\dagger c-d^\dagger d,
\end{equation}
where $c$ and $d$ are the annihilation operators of the output
beams. They can be expressed (taking into account the additional
phase shift of $90^{\circ}$) in terms of the annihilation operators
of the radiation inside the interferometer:
\begin{equation}
c=\frac i{\sqrt{2}}(a_1+a_2)\qquad\text{and}\qquad
d=\frac1{\sqrt{2}}(-a_1+a_2),
\end{equation}
where $a_1$ and $a_2$ act on states in the first and the second arm,
respectively. Thus, in terms of the operators $a_1$ and $a_2$, the
observable \eqref{charge} equals
\begin{equation}\label{obs1}
\widehat{I}=\ak_1\otimes a_2+a_1\otimes\ak_2.
\end{equation}
By definition of $a$ and $\ak$,
\[a\,|n\ket=\sqrt{n}\,|n-1\ket,\quad\ak\,|n\ket=\sqrt{n+1}\,|n+1\ket,\]
we obtain
the expectation of the observable \eqref{obs1} and its variance:
\begin{gather}
I(t)=\tr\{\widehat{I}\rho_t\},\qquad D(t)=\tr\{\widehat{I^2}\rho_t\}-I^2(t), \label{tr}\\
\widehat{I^2}=(\ak_1)^2\otimes(a_2)^2+2\ak_1a_1\otimes\ak_2a_2+\ak_2a_2+\ak_1a_1+(a_1)^2\otimes(\ak_2)^2,\label{meansq2}
\end{gather}
where $\rho_t=\rho_1(t)\otimes\rho_2(t)$.
$I(t)$ and $D(t)$ are calculated  explicitly:
\begin{align}
I(t)&=2 N \sigma \sqrt{1 - \sigma^2}e^{-g^2 c_t-2 N \sigma^2
\sin^2(g^2 s_t)} \sin \bigl\{g(2\varphi_t + g s_t) + N \sigma^2
\sin (2 g^2 s_t)\bigr\},\label{mean2}\\
D(t)&=N + 2N^2\sigma^2 (1 - \sigma^2)-2 N^2 \sigma^2 (1 -
\sigma^2)e^{-4 g^2 c_t -2N \sigma^2 \sin^2(2g^2 s_t)} \notag\\
&\times\cos \bigl\{4 g(\varphi_t + g s_t)  + N\sigma^2\sin
(4g^2s_t)\bigr\}-I^2(t),\label{disp2}
\end{align}
where
\begin{align}
\varphi_t &= \int_0^t\int_0^\tau d\tau\,ds \,f(s)e^{\frac\lambda
2(s-\tau)}\sin\Omega(s-\tau),\label{force}\\
c_t&= \int_0^t\int_0^\tau d\tau\,ds \,e^{\frac\lambda
2(s-\tau)}\cos\Omega(s-\tau)\notag\\
&\;=\frac{\frac{-\lambda^2}{4} + \frac{\lambda^3 t}{8} + \Omega^2 +
\frac{\lambda t\Omega^2}2} {\left( \frac{\lambda^2}4 + \Omega^2
\right)^2} + e^{\frac{-\lambda t }2}\,\frac{\cos\left(\Omega\,t+\phi\right)}{ \frac{\lambda^2}4 +\Omega^2 }, \label{cos}\\
s_t&=\int_0^t\int_0^\tau d\tau\,ds \,e^{\frac\lambda
2(s-\tau)}\sin\Omega(s-\tau)\notag\\
&\;=-\Omega\,\frac{ \frac{\lambda^2 t}4 + t\Omega^2-\lambda
}{\left(\frac{\lambda^2}4 +\Omega^2 \right)^2} - e^{-\frac{\lambda
t}2}\, \frac{\sin\left(\Omega\,t+\phi\right)}{\frac{\lambda^2}4
+\Omega^2}, \label{sin}\\
\tan\phi&=\frac{\lambda\Omega}{\frac{\lambda^2}4-\Omega^2}.\notag
\end{align}

The output signal \eqref{mean2} depends on
the classical force $f(t)$ and the pressure
of the radiation in the coherent state. If
the aim is to detect the
classical force, one should single out the
function $\varphi_t$. To this end,
according to
\eqref{mean2}, it is necessary to decrease
the reflectivity $\sigma$ of the first
splitter and, in this way, to reduce the
influence of the laser beam on
the oscillator.

Alternatively, the first beam splitter is taken with a splitting
ratio 50/50 ($\sigma=1/\sqrt2$), but the second arm should be a
cavity with a movable mirror like that in the first one. Then, as we
will see below, in the first approximation, the output signal $I(t)$
will depend only on $\varphi_t$.

Let the phase of the external force in the second cavity be opposite
to that in the first cavity. Then the maximum sensitivity of the
device will be attained. Therefore, the state $\rho_2$ of the
radiation in the second cavity is given by \eqref{rhol} but with
``minus'' sigh at $f(t)$:
\begin{align}
\rho_1(t) &= e^{-\frac N 2}\sum_{n=0}^\infty \sum_{m=0}^\infty
|n\ket_1 \phantom{i}_1\bra m| \frac{(i z)^n (-i z^*)^m}{2^{n/2} 2^{m/2}\sqrt{n!m!}} e^{-i\omega(n-m)t}\notag\\
&\qquad\times e^{g^2(n-m)\int_0^t\int_0^\tau d\tau ds \,\left(m
\,e^{(-i \Omega +\frac\lambda2)(s-\tau)}-n \,e^{(i \Omega
+\frac\lambda2)(s-\tau)}\right)}\notag\\
&\qquad\times e^{-2 i g(n-m)\int_0^t\int_0^\tau d\tau ds \,f(s)
e^{\frac\lambda2(s-\tau)}\sin\Omega(s-\tau)}\,,\\
\rho_2(t) &= e^{-\frac N 2}\sum_{n=0}^\infty \sum_{m=0}^\infty
|n\ket_2 \phantom{i}_2\bra m| \frac{z^n  (z^*)^m}{2^{n/2} 2^{m/2}\sqrt{n!m!}} e^{-i\omega(n-m)t}\notag\\
&\qquad\times e^{g^2(n-m)\int_0^t\int_0^\tau d\tau ds \,\left(m
\,e^{(-i \Omega +\frac\lambda2)(s-\tau)}-n \,e^{(i \Omega
+\frac\lambda2)(s-\tau)}\right)}\notag\\
&\qquad\times e^{2 i g(n-m)\int_0^t\int_0^\tau d\tau ds \,f(s)
e^{\frac\lambda2(s-\tau)}\sin\Omega(s-\tau)}\,.
\end{align}
From \eqref{tr} and \eqref{meansq2},
we obtain
\begin{gather}\label{mean}
I(t) = N e^{-2 g^2 c_t -2 N \sin^2 (g^2 s_t)}\sin (4 g \varphi_t)
\approx 4 N g \varphi_t,\\
D(t) = N + \frac{N^2}2 - \frac{N^2}2e^{-8 g^2 c_t -2
N\sin^2(2g^2s_t)} \cos (8g \varphi_t)-I(t)^2\notag\\
\shoveright{\approx N + 4 N^2 g^2 c_t + 4 N^3 g^4 s_t^2}\label{disp}
\end{gather}
provided the force $g \varphi_ t \ll 1$ and the coupling constant
$g^2 c_t \ll 1$, $g^2 s_t \ll 1$ are small enough. However, as one
can see from \eqref{sin}, the function $|s_t|$ increases with time,
therefore in the domain
\begin{equation}\label{nmax}N g^4 s_t^2\gg1\end{equation} approximations
\eqref{mean}, \eqref{disp} are non-applicable, and $I(t) \rightarrow
0$ and $D(t) \rightarrow \frac{N^2}2$
exponentially with respect to
$N g^4 s_t^2$.

Below, we consider the condition
$$\sigma^2_t\le1,\quad
\sigma^2_t=\frac{D(t)}{I^2(t)}
$$
as a relevant signal-to-noise ratio which
allows one to detect the external force.

\section{Numerical results}
Let us estimate the sensitivity of an interferometer of the LIGO
type \cite{ligo} by using the above formulas. The external force
which should be detected is created by a gravitational wave acting
on two widely separated masses that are suspended mirrors in the
arms of interferometer. Suppose that the incident wave propagates
transversely to the planar interferometer and the force acting on
the mirrors is equal to $F(t)= F_m\sin (\omega_{gr}t+\phi)$, where
(see \cite{bra})
\[
F_m = h L m \omega_{gr}^2,
\]
$h$ is the dimensionless amplitude of metric perturbations,
$\omega_{gr}$ is the frequency of the gravitational wave. The
detector LIGO measures external forces in free masses mode, i.e. the
oscillator eigenfrequency $\Omega$ is much less than that of the
gravitational wave $\omega_{gr}$. The required parameters of this
interferometer take the following values (see \cite{ligo}):
\[
L=4\times10^3 \text{ m},\quad \omega \sim 10^{15} \text{
sec}^{-1},\quad m\sim10\text{ kg}, \quad\Omega \sim 2 \pi \text{
sec}^{-1}, \quad \omega_{gr}\sim 2\pi10^2\text{ sec}^{-1}.
\]
According to \cite{ligo}, the coupling constant \eqref{constg} and
the force amplitude \eqref{constf} take the following values:
\[
g=\frac{\omega}{L}\sqrt{\frac{2\hbar}{m\Omega}}\approx
8.1\times10^{-7}\text{ sec}^{-1},\qquad f_m
=\frac{F_m}{\sqrt{2m\Omega\hbar}}\approx1.37\times10^{26}(\text{sec}^{-1})\,h.
\]
As the damping parameter $\lambda$ of the suspended mirrors does not
exceed $10^{-5}\text{ sec}^{-1}$ \cite{ligo}, for reasonable values
of $t$, only leading terms of $c_t$, $s_t$, $\varphi_t$
\eqref{force}--\eqref{sin} can be preserved in expansions in
$\lambda$:
\begin{gather}
\varphi_t = \frac{f_m (\cos
   \Omega t-1)\cos \phi}{\Omega  \omega_{gr}}-\frac{f_m  \sin
    \Omega t\sin \phi}{\omega_{gr}^2}\,,\label{apfi}\\
c_t=\frac1{\Omega^2}-\frac{\cos \Omega t
   }{\Omega ^2},\qquad
s_t=\frac{\sin \Omega t}{\Omega
   ^2}-\frac{t}{\Omega }\,,\label{apsc}
\end{gather}
because $\Omega\ll\omega_{gr}$. The absolute
value of $\varphi_t$ strongly depends on the
initial phase $\phi$ of the force. For two
extreme values $\phi=0$ and
$\phi=\frac\pi2$, the ratio of the
amplitudes is
\[
\frac{\varphi_m(\phi=0)}{\varphi_m(\phi=\frac\pi2)}=\frac{\omega_{gr}}{\Omega}=100\,.
\]
Further, we consider the most favorable case: $\phi$ is close to
zero.

Let us estimate the amplitudes of the oscillating functions $c_t$,
$s_t$ and $\varphi_t$ \eqref{apsc}, \eqref{apfi}. If the duration of
the measurement is longer than the period $T=2\pi/\Omega$ of the
movable mirror, the amplitudes are
\begin{equation}\label{amp}
\varphi_m=\frac{2f_m}{\Omega\omega_{gr}},\quad c_m =
\frac2{\Omega^2}, \quad s_m= \frac t\Omega.
\end{equation}
For parameter values  of the LIGO detector,
approximations \eqref{mean},
\eqref{disp} are quite
accurate:
\begin{equation}\label{num}
g\varphi_m \approx 5.6\times10^{16}h,\quad g^2 c_m\approx3.3\times
10^{-14},\quad g^2s_m\approx10^{-13} (\text{sec}^{-1}) \,t,
\end{equation}
and the expected fluctuations $h$ of metric does not exceed
$10^{-20}$ \cite{bra}.

In accordance with \eqref{nmax}, we suppose that the following upper
bound for the photon number holds true:
\begin{equation}\label{nmaxt}
N\ll N_{m} = \frac{\Omega^2}{g^4 t^2
}\approx10^{26}\text{(sec}^2)\,t^{-2}.
\end{equation}

Let us characterize the
sensitivity of the detector by the
square of relative fluctuations
\begin{equation} \label{fluct}
\sigma^2(t) = \frac{D(t)}{I(t)^2} \approx
\frac1{16g^2\varphi_t^2N}+\frac{c_t}{4\varphi_t^2}+\frac{g^2
s_t^2N}{4\varphi_t^2}.
\end{equation}
The first summand is the leading term provided the number  $N$ of
photons is small. It corresponds to the Poisson fluctuations of the
photon number if the field in the cavity is in the coherent state.

The second term is the Heisenberg uncertainty relation for quantum
oscillator, and the third term characterizes the perturbation of the
oscillator dynamics by quantum noise of the laser field.

For definiteness, let us assume that detection of the external force
is possible if the relative fluctuation (\ref{fluct}) is less than
1. Substituting the amplitude values \eqref{amp} to \eqref{fluct},
we estimate the second term of the sum \eqref{fluct}:
\[
\frac{c_m}{4\varphi_m^2} \approx\frac{2.6\times10^{-48}}{h^2}<1.
\]
Thus, in the framework of this model, the detection of the
gravitational wave is impossible if $h<10^{-24}$ for any number of
photons in the cavity.

Further, we assume that $h\sim10^{-22}$, so the second term in
\eqref{fluct} can be omitted. Then the lower bound for the number of
photons is given by the first term of \eqref{fluct}:
\[
\frac1{16g^2\varphi_m^2N}<1, \quad \text{or}\quad  N>
N_{min}=2\times10^{9},
\]
that corresponds to the laser power
\[
P_{min}= \frac{\hbar\omega
N_{min}}{1000}\frac{c}{2L}\approx1.4\times10^{-8}\text{ w}
\]
under the assumption that the mean number of
beam reflections in the cavity is $10^3$.
The third summand determines the upper bound
for the number of photons:
\[
\frac{g^2 s_m^2 N}{4\varphi_m^2}<1, \quad \text{or}\quad N<N_{max}=
\frac{1.2\times10^{16}(\text{sec}^2)}{t^2}.
\]
Indeed, the inequality \eqref{nmaxt}
holds true for the above
magnitude $f_m$ of the gravitational force.

\begin{figure}
\centering
\includegraphics{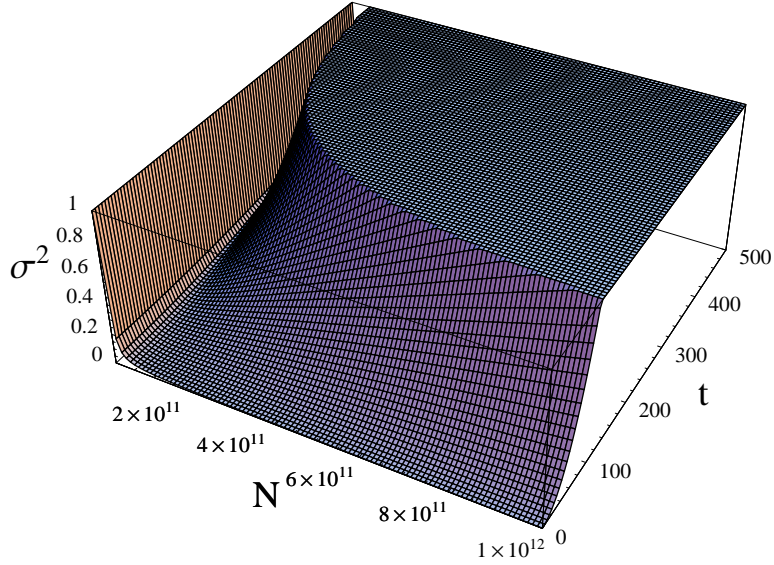}
\caption{The dependence of the
relative fluctuations on the measurement
time and the number of photons.}\label{gr_fluct}
\end{figure}

\begin{figure}
\centering
\includegraphics{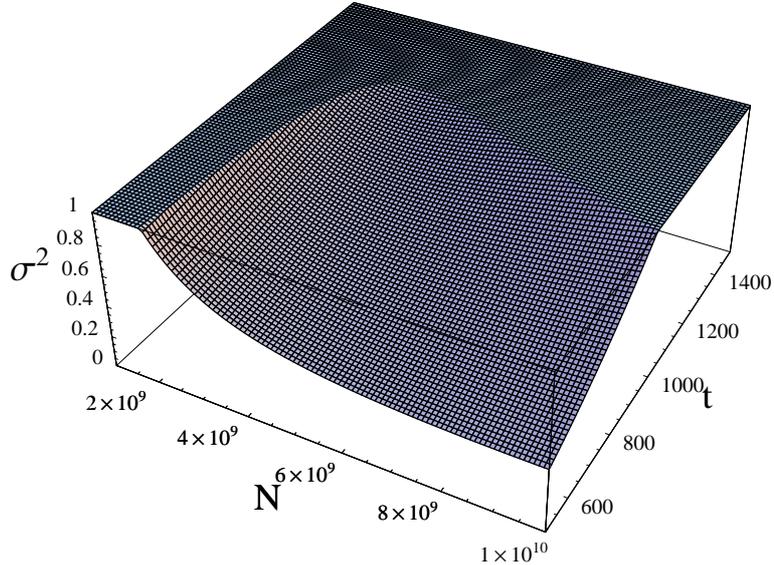}
\caption{The bounds for the maximum measurement
time.}\label{gr_fluct1}
\end{figure}

The maximum  number $N_{max}$ of photons
depends on the measurement time. In
particular,
\begin{align*}
N_{max} &= 10^{16},& P_{max}&=\frac{\hbar\omega
N_{max}}{1000}\frac{c}{2L}\approx8\times10^{-2}\text{ w},&&
\text{for }t=1\,\text{sec},\\
N_{max} &= 10^{12},& P_{max}&=\frac{\hbar\omega
N_{max}}{1000}\frac{c}{2L}\approx8\times10^{-6}\text{ w},&&
\text{for }t=100\,\text{sec}.
\end{align*}
If $t\sim1000$, the upper $N_{max}$ and lower $N_{min}$ bounds of
the number of photons attain each other and become inconsistent for
longer measurements. The typical dependence of the detector
sensitivity on the measurement time and the number of photons is
given in Fig. \ref{gr_fluct}. The cutoff of the graph is made at the
points where the relative fluctuation attains 1.
From \eqref{fluct}, one can find the optimal number of photons, i.e.
the point, where $\sigma^2(t)$ reaches minimum in  $N$:
\[
N_\text{opt} =
\frac1{2g^2s_m}\sim\frac{4.8\times10^{12}(\text{sec})}{t},\qquad
\sigma^2(N_\text{opt}) = \frac{s_m}{4\varphi_m^2}<1.
\]
Consequently, for $h\sim10^{-22}$,
we find the maximum
measurement duration:
\[
t_\text{max}=\frac{16f_m^2}{\Omega\omega_{gr}^2}\approx1200\text{
sec},
\]
as represented in Fig.
\ref{gr_fluct1}.

\begin{figure}
\centering
\includegraphics{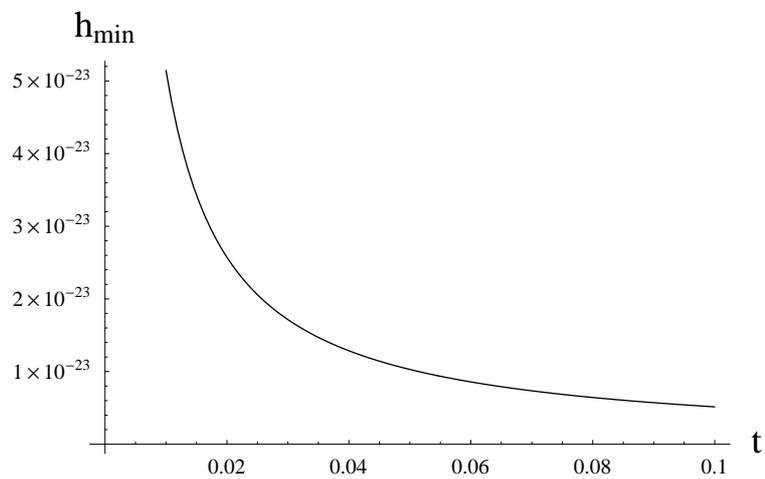}
\caption{The minimal detectable metric fluctuation.}\label{gr_metr}
\end{figure}

\begin{figure}
\centering
\includegraphics{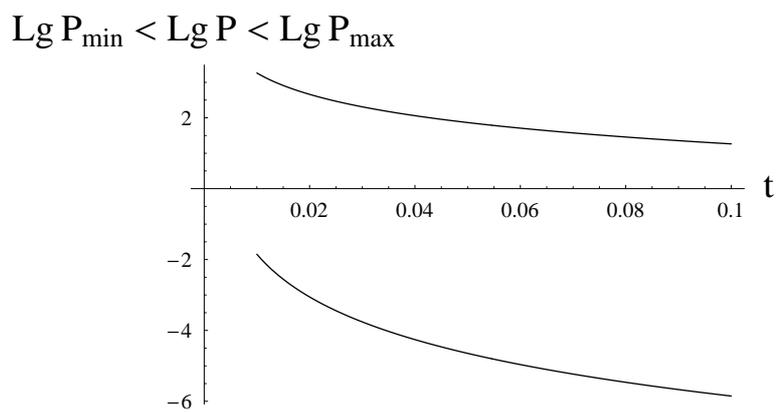}
\caption{The limitation on the laser
power.}\label{gr_p}
\end{figure}

Taking the time of measurement be equal to several periods of the
gravitational wave $t\sim0.01 \text{ sec}$, we obtain the following
approximate expressions for the functions $\varphi_t$, $c_t$ and
$s_t$:
\begin{equation}
|\varphi_t|\approx\frac{f_mt^2\Omega}{2\omega_{gr}},\quad
c_t\approx\frac{t^2}2,\quad |s_t|\approx\frac{t^3\Omega}6.
\end{equation}
In Figs. \ref{gr_metr} and \ref{gr_p}, we present the dependence of
the minimal detectable fluctuations of metric with regard to the
time of measurement and the bounds on the laser power which follow
from the restriction $\sigma^2\leq1$.
%\begin{figure}
%\centering
%\includegraphics{pmax.eps}
%\caption{Ограничение на максимальную мощность
%лазера.}\label{gr_pmax}
%\end{figure}

\section{Conclusion}
Let us summarize briefly the main results of our study. First,
formulas \eqref{mean2}, \eqref{disp2} and \eqref{mean}, \eqref{disp}
give explicit expressions for the signal and its variance on the
output of a two-arm interferometer measuring small classical forces.
These formulas take into account full quantum description of system
dynamics and irreversible transfer of the system energy to the
environment at zero temperature. We point out a fast decrease of the
signal and the convergence of its variance to a constant, so that
the relative fluctuations tend to infinity \eqref{nmax}. Moreover,
as it follows from \eqref{fluct}, the main quantum  noises appear as
leading terms in the expansion of the relative fluctuations with
respect to the small coupling constant $g^2$; they are (i) the
Poisson fluctuations of the number of photons (the shot noise), (ii)
the standard quantum limit uncertainty relation (due to CCR), (iii)
quantum coupling between the oscillator and the laser radiation (the
light pressure).

Explicitly calculated density matrix \eqref{rho} of the system (the
radiation $\otimes$ the oscillator) allows one to find the mean
values of observables and their variances for arbitrary initial
states of the system (e.g. squeezed states).

The study of the oscillator interacting with the electromagnetic
radiation, a classical force and the environment at nonzero
temperature described by the generator \eqref{hibbs} is a more
difficult but quite relevant problem. Our approach  will be
presented in the future paper \cite{npub}.

The author acknowledges Prof. V.P.
Mitrofanov for his helpful discussions.

\end{document}